\documentclass[conference]{IEEEtran}
 \pdfoutput=1 
 
\ifCLASSINFOpdf
   \usepackage[pdftex]{graphicx}
   \graphicspath{{../pdf/}{../pdffig/}}
   \DeclareGraphicsExtensions{.pdf,.jpeg,.png}
\else
   \usepackage[dvips]{graphicx}
\fi
\usepackage{multirow}
\usepackage{breakurl}
\usepackage[breaklinks]{hyperref}
\usepackage{flushend}

\def \main {\texttt{main} }
\def \AsyncTask {\texttt{AsyncTask} }

\newcommand{\tabincell}[2]{\begin{tabular}{@{}#1@{}}#2\end{tabular}}

\begin{document}
%no one aware
\title{PersisDroid: Android Performance Diagnosis via Anatomizing Asynchronous Executions}

\author{\IEEEauthorblockN{Yu Kang, Yangfan Zhou, Hui Xu, and Michael R. Lyu}
\IEEEauthorblockA{Dept. of Computer Science \& Engineering, The Chinese Univ. of Hong Kong,
Hong Kong\\ Email: \{ykang, yfzhou, hxu, lyu\}@cse.cuhk.edu.hk}}

\maketitle

\begin{abstract}

Android applications (apps) grow dramatically in recent years.
Apps are user interface (UI) centric typically.
Rapid UI responsiveness is key consideration to app
developers. However, we still lack a handy tool for
profiling app performance so as to diagnose performance problems.
This paper presents \texttt{PersisDroid}, a tool specifically designed for this task.
The key notion of \texttt{PersisDroid} is
that the UI-triggered asynchronous executions also contribute to the UI performance, and hence
its performance should be properly captured to facilitate performance diagnosis.
However, Android allows tremendous ways to start the asynchronous executions, posing a great
challenge to profiling such execution.
This paper finds that they can be grouped into six categories. As a result, they can be tracked
and profiled according to the specifics of each category with a dynamic instrumentation approach
carefully tailored for Android.
\texttt{PersisDroid} can then properly profile the asynchronous executions in task granularity, which
equips it with low-overhead and high compatibility merits. Most importantly, the profiling data
can greatly help the developers in detecting and locating performance anomalies.
We code and open-source release \texttt{PersisDroid}. The tool is applied in diagnosing 30 open-source
apps, and we find 20 of them contain potential performance problems, which shows its effectiveness
in performance diagnosis for Android apps.
\end{abstract}

\section{Introduction}
Smartphones have surged into popularity in recent years, which have changed the way people live, work and have fun.  As a personal device for daily life, smartphone applications (apps) are expected to provide a quick response to the input from the user interface (UI).  UI performance is a critical factor to user experiences.

However, poor UI responsiveness of many Android apps is still a widely-complaint type of bugs  \cite{AndroidPerf:CharacterizingBugs}.  Without a suitable tool, it is not easy for app
developers to understand how their apps perform and according conduct performance diagnosis.
The current practice of debugging Android performance issues
generally resorts to Android method tracing tools ({\em e.g.,} Traceview and dmtracedump \cite{AndroidPerf:MethodTracing}). However, these tools require daunting human efforts in analyzing the tremendous data they produce. A performance diagnosis tool that can facilitate the detecting and locating of performance problem is still at large.

This paper presents the \texttt{PersisDroid} tool we specifically designed for this task.
Unlike existing tools that focus on detecting performance problems in the UI thread ({\em e.g.},
StrictMode \cite{AndroidPerf:StrictMode} and Asynchronizer \cite{AndroidPerf:RefactorAsynctask}),
the \texttt{PersisDroid} design is motivated by the fact that the UI-triggered asynchronous
executions in other threads also contribute to the UI performance. We therefore propose to capture such asynchronous executions and properly profile their performance. The performance data can then be
analyzed for effective performance diagnosis purpose.

Unfortunately, capturing and profiling the asynchronous executions is very challenging. The
difficulties include: 1) There are tremendous ways to start the asynchronous executions. As a result, there
exist no single entry and exit points for such executions. It is not easy to track them. 2) A general way to capture and profile such execution is via instrumentation, {\em i.e.}, tracking
the execution during app runtime. However, there are a large number of smartphone models available. Different manufacturers have their own customized Android OS kernels and frameworks. The tool should be
compatible to various smartphone models, which cannot resort to kernel or framework recompilation since
their source codes are typically unavailable for most smartphone models. 3) The tool should not incur
high overhead to the app execution so as to guarantee the efficiency of performance diagnosis task.

To solve these challenges, we first observe that the tremendous ways to start the asynchronous executions
can be actually grouped into six categories.
\texttt{PersisDroid} can then track
and profile the asynchronous executions
for each category in task granularity, according to the specifics of each category.
\texttt{PersisDroid} solves the compatibility and efficiency challenges by slightly instrumenting only
the general framework methods with a dynamic instrumentation approach.
As a result, \texttt{PersisDroid}
can be applied to most available smartphone models in the market to capture and profile the asynchronous executions efficiently. Most importantly,  \texttt{PersisDroid} can rely on the profiling data in greatly helping the developers in detecting and locating performance anomalies.

We code and open-source release \texttt{PersisDroid} at \cite{AndroidPerf:PersisDroid}. We apply the tool
to performance diagnosis tasks for 30 open-source apps. We find 20 of them contain potential performance problems. Our experiences demonstrate
the effectiveness of \texttt{PersisDroid} in performance diagnosis for Android apps.

The rest of the paper is organized as follows. Section II provides the preliminary knowledge. In Section III, we show the motivating
examples of performance problems that are caused by synchronous executions. Section IV overviews \texttt{PersisDroid}, with its details
discussed in Section V and VI. Section VII provides our experimental study, and the paper is concluded in Section VIII.

\section{Android Application Specifics}
\label{paper_android_perf_icse: specifics}

Designed mainly for user-centric usage pattern, Android apps are typically user-interface (UI) oriented: A typical app will iteratively process user inputs, and accordingly update the display to show the intended contents. UI components are the building blocks of the UI. An representative UI component is the Android \texttt{View} \cite{Android:View}, which includes, for example, the \texttt{TextView} objects to display text, \texttt{ImageView} objects to display images, and \texttt{Button} objects. \texttt{Activity} \cite{Android:Activity} object is that to provides a window container for the UI components, shown in the display to the user.

When an app is launched, a unique \main thread will be created for the app. The \main thread is the sole thread that takes care of UI-related operations, such as processing user inputs and displaying the UI components ({\em e.g.}, buttons and image views). When a valid user input ({\em i.e.}, a UI event) comes, the \main thread can invoke its corresponding {\em UI event handler}, {\em i.e.}, the codes that processing the UI event accordingly.

Some UI event handlers may be time-consuming, for example, one that requires to download a large file from the Internet. To avoid blocking the \main thread, as suggested by the official Android development guide \cite{AndroidPerf:Threads}, UI event handler should start some concurrent executions ({\em e.g.}, by creating a new worker thread) for completing such heavy-weighted tasks in an asynchronous manner so that the \main thread can handle other UI inputs simultaneously.  After the asynchronous part of the UI event handler is done, the UI can also be updated in \main thread with a call-back mechanism.

\begin{figure}[!t]
\centering
\includegraphics[width=0.48\textwidth]{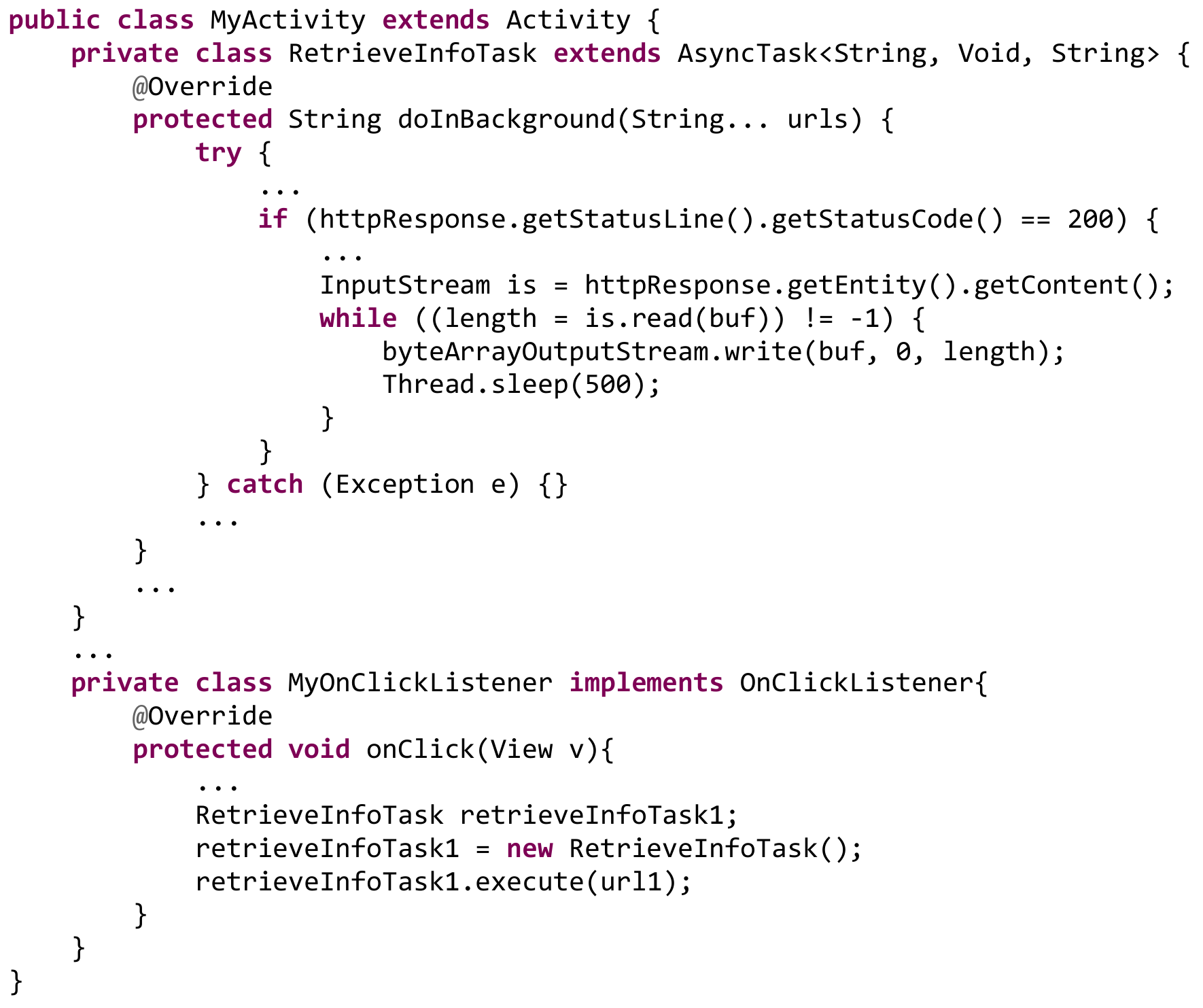}
\caption{An example of using AsyncTask}
\label{fig:perf_test_original_asynctask}
\end{figure}

Figure \ref{fig:perf_test_original_asynctask} shows the codes of an \texttt{Activity} which can retrieve some data from the Internet and display the data in a \texttt{TextView} after the user touch a button. The Internet access is done asynchronously in another thread while the \texttt{TextView} update is done in \main thread. More specifically, the \texttt{RetrieveInfoTask} extends the \texttt{AsyncTask} class. It overrides the \texttt{doInBackground} method to allow to access the Internet asynchronously in another thread. Its \texttt{onPostExecute} methods is a call-back mechanism to allow the \main thread to update a \texttt{TextView} object according.
These codes are abstracted from an open source project, namely, RestC \cite{AndroidPerf:RestC}. It shows a common coding practice for Android apps.

Android provides developers high flexibility to implement asynchronous code segments. There are tremendous ways for an app to start asynchronous executions. Examples include using \texttt{AsyncTask}, \texttt{ThreadPoolExecutor}, and \texttt{IntentService}. Actually we can find that hundreds of classes or methods in Android framework can initialize asynchronous executions, by inspecting the source codes of the Android framework \cite{Android:AndroidFrameworkSource}.

Implicit ways to start asynchronous executions include, for example, via the new classes that overrides the Android framework classes such as \texttt{Thread} or \texttt{AsyncTask}. Sometimes, it is even difficult for the developer herself to be aware that the codes she writes would start asynchronous executions. For example, when conducting a database operation, one may call \texttt{db.execSQL(...)}, which is actually implemented with asynchronous execution by the Android framework. Another example is when using the the text-to-speech service, the code like \texttt{textToSpeech.speak(...)} is implemented with asynchronous execution.

The complex ways of starting asynchronous executions makes it difficult for the developers to comprehend how the UI event handlers they write perform.  Performance problems are therefore hard to be avoided.  Next we will show some motivating examples of such code defects and how call stack as well as time of every phase of asynchronous execution is recorded and utilized for diagnosing a performance issues.

\section{Android Performance Problems: Motivating Examples}\label{sec:perf_test_perfproblems}

%UI-handling is centric to Android apps in general. When a user performs a valid UI %operation ({\em e.g.}, touching a button in the touchscreen) in an app, the operation is %handled as a UI event by the corresponding UI event handler. The handler can then conduct %the intended display update after processing the event.

Since there is only one sole \main UI thread for each app, UI events are handled one by one by the thread. Hence, in order not to block the \main thread, the Android official Android development guide \cite{AndroidPerf:Threads} suggests that asynchronous executions should be conducted for time-consuming tasks if an event handler include such tasks.

Actually app developers may overlook such suggestions occasionally and write time-consuming codes in the synchronous part of the UI event handler ({\em i.e.}, that executes in the \main thread). This is long been known as a notorious cause of performance problems \cite{AndroidPerf:Threads}, including frequent ANR (Application Not Responding) reports which indicate the consecutive UI events are blocked for more than $10$ seconds \cite{AndroidPerf:ANR}. Many tools have been designed to find such bugs. Examples include the official Google developer tool StrictMode \cite{AndroidPerf:StrictMode} and Asynchronizer \cite{AndroidPerf:RefactorAsynctask}, which typically address this issue via static analysis.

However, addressing the performance problems solely in the synchronous executions is far from enough. When a user thinks that she is suffering from slow and laggy UI, she is actually experiencing a long period of time between her UI operation and its corresponding intended display update. Event if asynchronous executions are introduced and the synchronous part can completes quickly, she may still feels that the UI is laggy if the asynchronous executions are slow and consequently cause the late display update of the intended contents.

Since Android allows complex ways in starting asynchronous executions, they may introduce various tricky performance problems. Next we will show even if simple, seemly-correct codes may contain performance problems.

We adopt \texttt{AsyncTask} as an example.  \texttt{AsyncTask} is a simple, handy class that allows developers to conveniently start self-defined asynchronous executions and notify the \main thread to update the UI \cite{AndroidPerf:Asynctask}.

We show two typical cases where performance problems are introduced. The first is caused by unexpected sequentialized asynchronous executions, while the second by not or not properly canceling expierd asynchronous executions.

\subsection{Sequential running of multiple asynchronous executions}
\label{subsec:perf_test_sequential}
\begin{figure}[!t]
\centering
\includegraphics[width=0.48\textwidth]{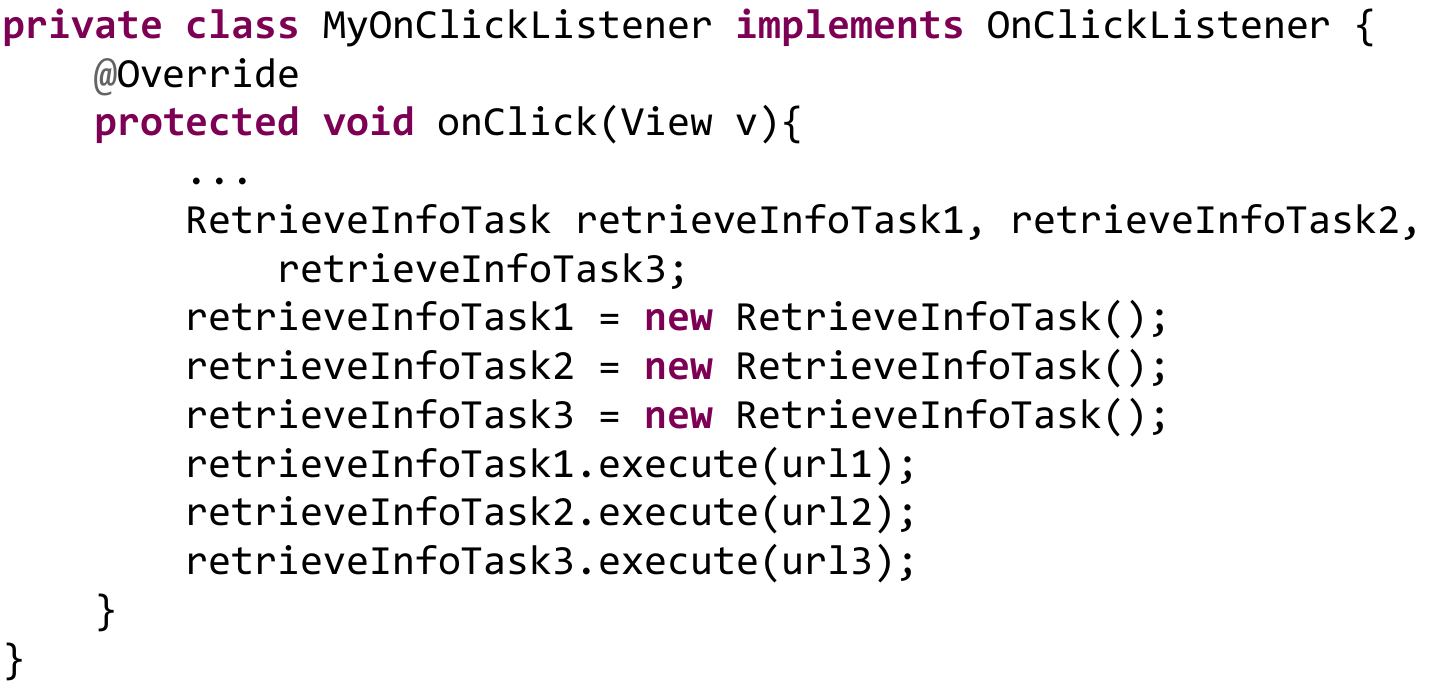}
\caption{Example codes which may cause potential performance problems}
\label{fig:perf_test_asynctask_pool_1}
\end{figure}

Let us suppose that an event handler will show three text views in an \texttt{Activity}.  The content of each text view should be loaded from the Internet.  Since Internet access is slow, the developer may resort to asynchronous executions to download the contents, and expect to download the three in parallel. Her codes are shown in Figure \ref{fig:perf_test_asynctask_pool_1}, where the class \texttt{RetrieveInfoTask} is defined in our previous example shown in Figure \ref{fig:perf_test_original_asynctask}.

Using the \texttt{execute} method is shown as a usage example by the official guide \cite{AndroidPerf:Asynctask}, which is commonly used to start the asynchronous executions defined by the \texttt{AsyncTask} extensions ({\em e.g.}, the \texttt{RetrieveInfoTask} class in this example).  The developer may consider that every \texttt{RetrieveInfoTask}s would be executed in separated threads, and hence calling their \texttt{execute} method in sequential will make them run in parallel.

However, even such simple codes contain a subtle cause of potential performance problem. Since the \texttt{execute} method of \texttt{AsyncTask} cannot be overridden, all three execution methods in this example will actually call the \texttt{execute} method implemented in their super class ({\em i.e.}, \texttt{AsyncTask}). In the recent versions of Android, invoking the \texttt{execute} method will insert the corresponding task into a global queue and all tasks will execute in sequential in one sole thread instead of in parallel in multiple threads. This inevitably incurs more time to complete the download tasks and update the UI accordingly. As a result, the user will experience slow UI.

It is worth noting that such sequential execution mechanism is introduced recently in Android systems with version numbers larger than 3.0. In previous versions, the codes will, in contrast, execute in parallel as expected. It is very possible for the developers to neglect such changes and introduce potential performance problems.  But current tools like StrictMode \cite{AndroidPerf:StrictMode} and Asynchronizer \cite{AndroidPerf:RefactorAsynctask} only locate problems in the synchronous executions.  As a result, such performance problems in the asynchronous executions caused by unexpected sequentialized asynchronous executions cannot be located conveniently by current tools. But, such code defects are quite common among developers. For example, the\cite{AndroidPerf:FacebookWidget} project contains a similar issue in its class \texttt{StreamFragment.StatusAdapter}.

Note that the execution time values of the tasks {\em per se} between the sequential case 
and the parallel case may not be quite different. 
However, for the sequential case, a task may be queuing for execution for a longer time after it is scheduled. 
If a tool can capture such a queuing time, it can greatly facilitate performance diagnosis. 
\texttt{PersisDroid} is a tool that can well capture such queuing time. As a result, it is able to detect 
and locate such performance problem. 
 
\begin{figure}[!t]
\centering
\includegraphics[width=0.5\textwidth]{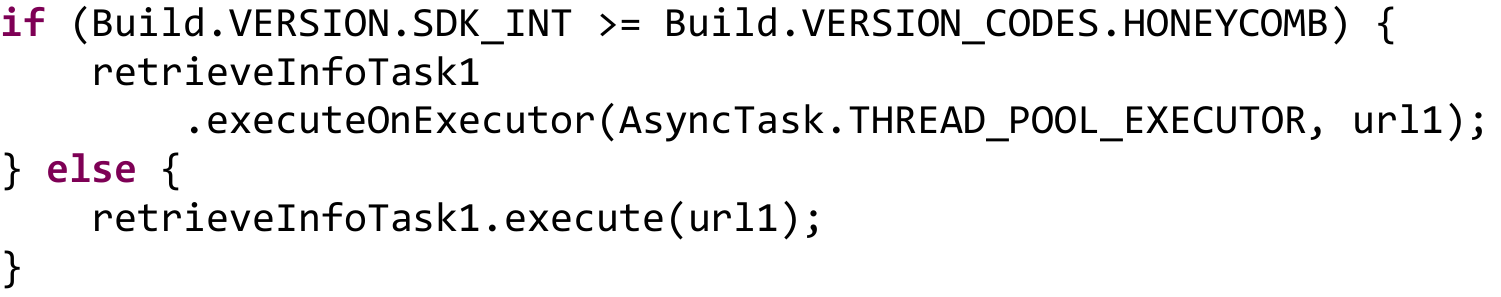}
\caption{Correct codes to execute an \texttt{Asynctask}}
\label{fig:perf_test_asynctask_pool_2}
\end{figure}

Finally, such a code defect can be resolved by invoking the \texttt{executeOnExecutor} method instead of the \texttt{execute} method by assigning a new task queue for each download task. We show the modifications of the first statement in Figure \ref{fig:perf_test_asynctask_pool_1} as an example in Figure \ref{fig:perf_test_asynctask_pool_2}.

\subsection{Not Canceling Unnecessary Asynchronous Executions}
\label{subsec:perf_test_nocancel}
Let us again consider the above example codes. After the codes are modified as shown in Figure \ref{fig:perf_test_asynctask_pool_2} so that the three tasks can execute in parallel, the codes still may cause performance problems. Suppose the app allows its user to switch the activities during the three views are being loaded in the current activity, for instance, with a right-to-left sliding operation. Such a mechanism is commonly used in Android apps, which facilitates users to quickly locate the activity of interest. An example is that an email client may allow the user to switch from the list-email activity to the read-email activity, even if the email list is not completely shown in the list-email activity.

When the user switches to another activity, the three views that the tasks are loading from the Internet are no longer required since their associated activity is invisible. It is therefore unnecessary to continue the three asynchronous tasks in downloading the Internet contents. But the example codes do not explicitly cancel the asynchronous tasks. The tasks can then run until the entire Internet contents are downloaded. Such unnecessary asynchronous executions may incur resource races ({\em e.g.}, occupying the Internet bandwidth), and therefore deteriorate the performance of other asynchronous executions. Sometimes, they may even block other asynchronous executions: For example, they can occupy the working threads in a thread pool and cause other tasks to be waiting for free threads.

\begin{figure}[!t]
\centering
\includegraphics[width=0.46\textwidth]{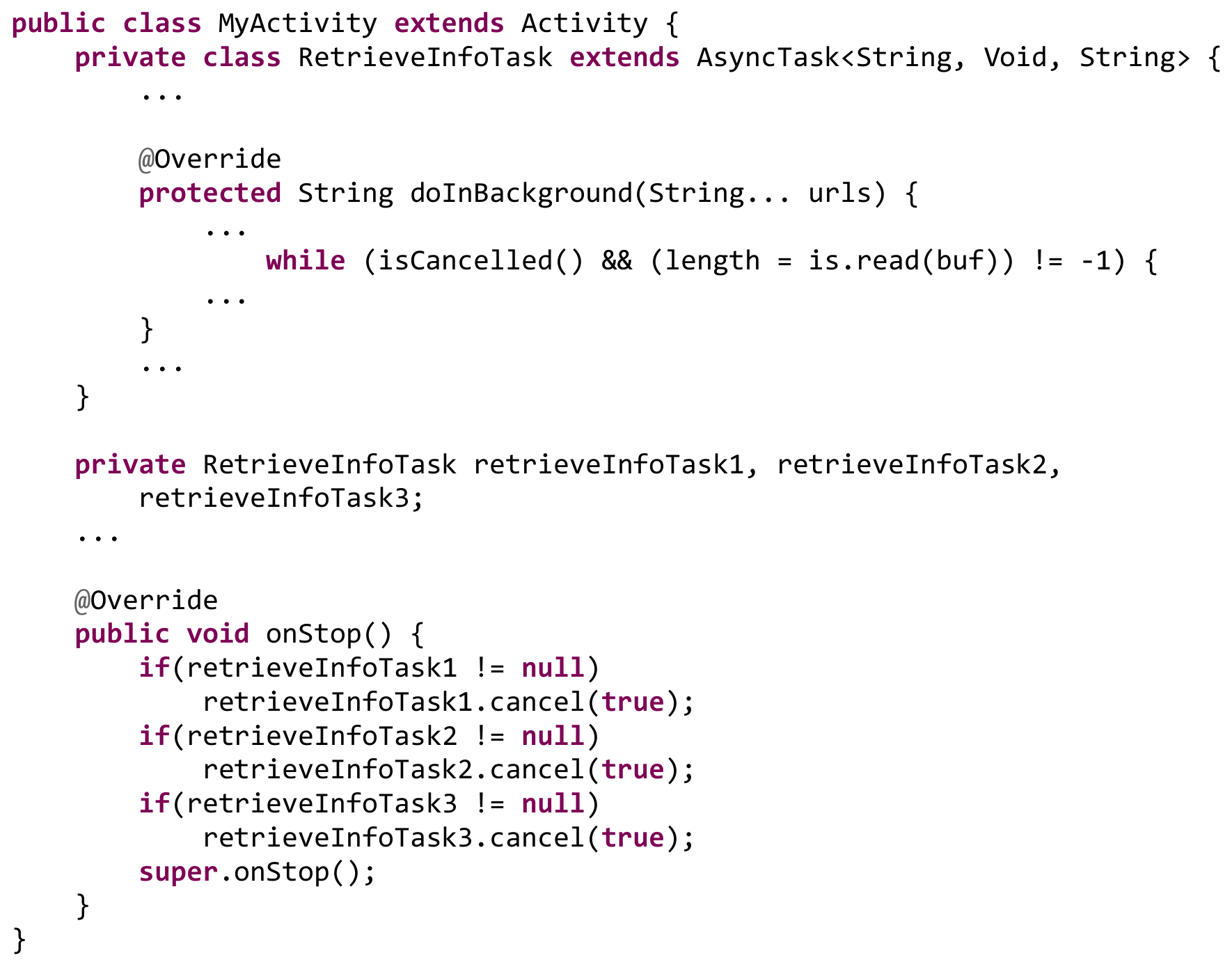}
\caption{An example of cancelling AsyncTask}
\label{fig:perf_test_asynctask_cancel}
\end{figure}

The codes can be further improved, as shown in Figure \ref{fig:perf_test_asynctask_cancel}.  Note that such code defects are very common to Android apps. We will show in our experimental study that many developers of the popular Android apps ({\em e.g.}, rtweel \cite{AndroidPerf:exampleproj_cancelonly} and BeerMusicDroid \cite{AndroidPerf:exampleproj_cancelvariable}) have made such mistakes.

\label{subsubsec:perf_test_no_cancel_example}

Note that a canceling operation is typically required when the task is time-consuming and should be terminated 
in the middle. If not properly canceling such a task, its execution latency will be relatively large. Hence, 
the key to detect and locate such a performance issue is to know the execution latency of the asynchronous tasks.
\texttt{PersisDroid} can well profile the task with its execution latency, which can greatly facilitate the diagnosis 
of such a performance problem.

Next, we will illustrate how \texttt{PersisDroid} detects and locates such subtle performance problems in asynchronous executions.

\section{Android Performance Problem Diagnosis: An Overview}
\label{paper_android_perf_icse: overview}

%: \texttt{Per}formance Diagno\texttt{sis} for An\texttt{droid} Apps
%Language                     files          blank        comment           code
%Java                            29            388            845           2506
%Python                          14            213            249           1212
%SUM:                            43            601           1094           3718
 To facilitate the debugging of performance defects of Android apps, we specifically design the \texttt{PersisDroid} tool.
The tool is implemented in Java and Python, which is open-source available at \cite{AndroidPerf:PersisDroid}. The current version contains over about $5000$ lines of codes, which cumulate more than 6 months of coding efforts.

\begin{figure}[!t]
\centering
\includegraphics[width=0.48\textwidth]{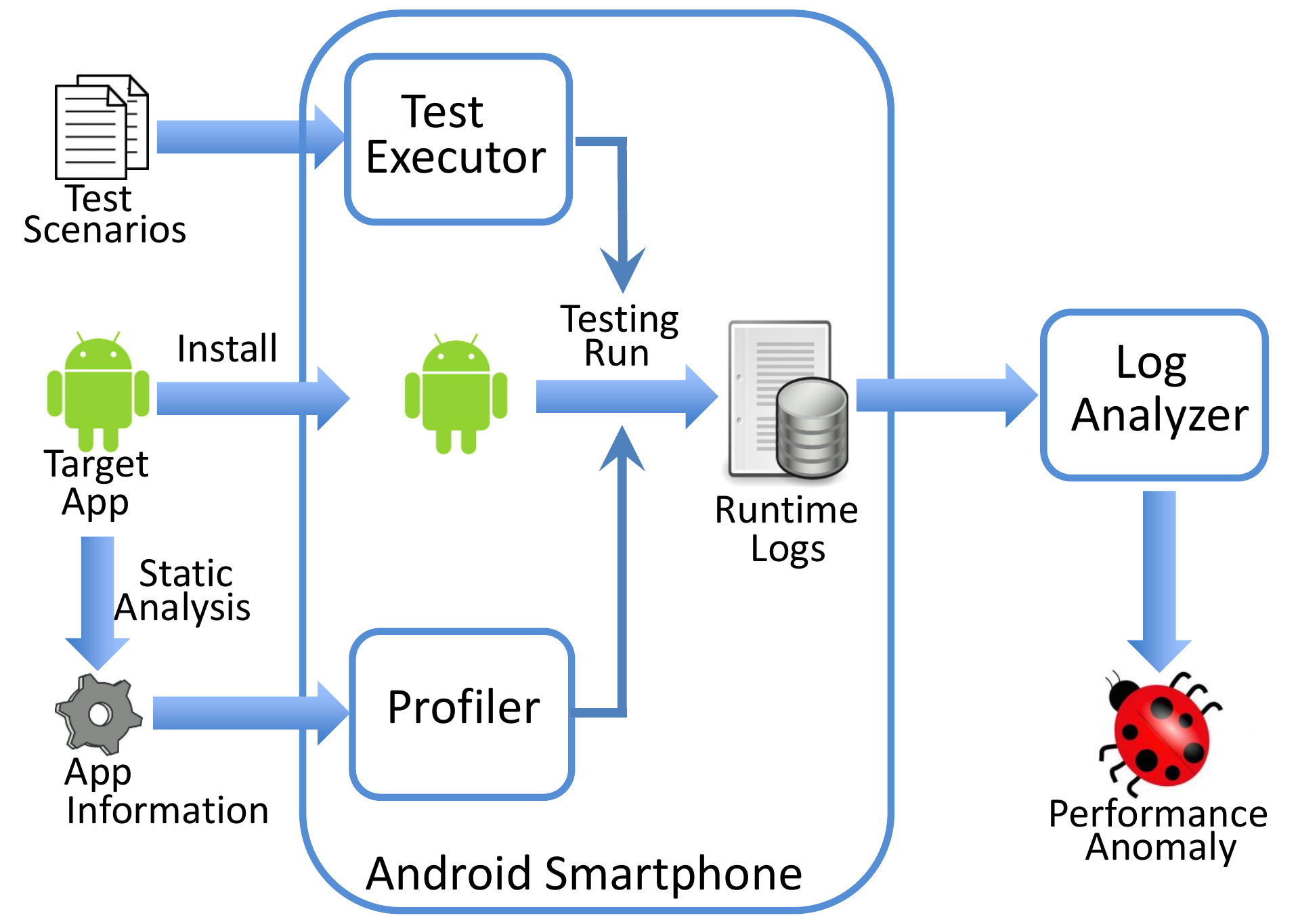}%{perf_test_framework}
\caption{Overview of \texttt{PersisDroid} Framework}
\label{fig:perf_test_framework}
\end{figure}

The framework of \texttt{PersisDroid} is shown in Figure \ref{fig:perf_test_framework}. It contains two parts: one runs on a computer and the other runs on a smartphone connected to the PC via a USB cable, which is shown inside the largest frame in the figure.

First, the target app is analyzed by the {\em static analysis} module so as to generate some required app information for the {\em profiler} module. \texttt{PersisDroid} then installs the app into the smartphone. It runs the app and starts a {\em test executor} to generate test cases ({\em i.e.}, typically user touchscreen inputs) for the app. The profiler module then captures the runtime logs of the app. The logs are then
pulled to the PC. A {\em log analyzer} can detect and locate performance anomalies. A generated report can finally direct the developer to inspecting the source codes of the app.

\texttt{PersisDroid} resorts to dynamic analysis to detect and locate performance
anomalies. To exercise the target app and trigger the potential performance problems, we need to adopt a testing tool in conducting the test runs for the app. \texttt{PersisDroid} does not requires a specific testing tool. Instead, we implement the test executor as a plugin, which can instantly adopt the existing, mature testing frameworks, including Monkey \cite{AndroidTest:Monkey})
for random testing or prevalent semi-automatic script testing tools like Robotium \cite{AndroidTest:Robotium} and UIAutomator \cite{AndroidTest:UIAutomator}. Even manual testing can be used to exercise the target app in \texttt{PersisDroid}.

One important design aim of \texttt{PersisDroid} is to detect and locate performance problems of Android apps in their
asynchronous executions. To this end, we first requires the profiler component of \texttt{PersisDroid} can identify
such executions during app runtime and properly profile them into a set of feature data. Given the tremendous ways
in starting asynchronous executions, this becomes a challenging task. The detailed design considerations of the profiler
will be discussed in Section \ref{paper_android_perf_icse: profile}.

With the data captured by the profiler ({\em i.e.}, the logs of the app runtime), we
then carefully tailor a log analyzer to eventually detect performance problems and reason their root causes. We will
illustrate the design of the log analyze in Section \ref{paper_android_perf_icse: analyzer}.

Finally, to make \texttt{PersisDroid} a practical, handy tool for performance problem diagnosis, we also need to carefully
consider the  system design requirements like {\em compatibility}, {\em usability}, and {\em low overhead}.
Specifically, \texttt{PersisDroid} should work for most mainstream Android smartphones, without
a requirement to the manufacturer specifics. In other words, it would rely on general features of mainstream Android
smartphones. Second, the tool should be convenient to use. Simple installation operations should be enough for the tool to
be working on a target smartphone. It must not require to recompile the operation system (OS) kernel or the Android framework,
since these operations in general incur daunting engineering efforts, not to mention their feasibility on diverse
smartphones from highly self-protected manufacturers. Finally, the tool cannot rely on heavy-weighted mechanisms. For example, it cannot requires
a heavy instrumentation to the target app. This is important for efficiency consideration.

\section{Profiling asynchronous executions}
\label{paper_android_perf_icse: profile}

We have shown that centric to diagnose the performance problems of Android apps is to profiling the execution of the asynchronous executions. In essential, the following problem should be carefully addressed in profiling the asynchronous executions for \texttt{PersisDroid}: What kind of data should be profile and how. This section discusses these key issues.

\subsection{Profiling Granularity and Feature Selection}

On one hand, the profiling granularity should be fine-grained enough so that it
not only should help detect a performance problem, but can direct human inspection of the codes so as to locate the defect that causes a performance problem. On the other hand, to achieve the required profiling granularity, the profiler must not incur much overhead, and bring usability issues, {\em i.e.}, the profiler should be compatible to most existing Android smartphones and easy to be applied to diverse apps in app market.

To this end, we suggest to profile asynchronous execution in {\em task} level, where a task is the developer-defined intended job that are assigned to run in an asynchronous manner either explicitly or implicitly. For example,
the codes in \texttt{RetrieveInfoTask.doInBackground} in Figure \ref{fig:perf_test_original_asynctask} define such a task.

First, we find that it is good enough in performance diagnosis to profile in such a granularity than in a finer one like in the line level or in the method level. This is because a task is a developer-intended execution with typically short codes: If the developer can know the performance of which task is anomalous, she can instantly reason its cause by inspecting the codes she write for the task. Secondly, such a granularity will not incur too much overhead to the profiler, comparing with the finer granularities ({\em e.g.}, in method level or in line level).

After determining the granularity, we then discuss the data we should profile for each task. We profile three features for each task. Specifically, the {\em execution context} of the task, the {\em execution latency} of the task, and how long it waits for execution, {\em i.e.}, the {\em queuing time} between it is scheduled and it starts.

The first feature is collected for identifying the task and facilitating to locate its source codes. We consider the call-stack where the app requests to start an asynchronous task as the context. A call-stack encapsulates the method invocation information together with the class names and method names, which can instantly help the developer in identifying the location of the task and where it is requested in the source codes. Moreover, it is easy obtain the call-stack with one line of java code: \texttt{Thread.currentThread().getStackTrace()}.

The latter two features are for detecting and locating performance anomalies. Execution latency is a straightforward feature that describes how fast the task runs. It is worth noting that we in addition take the queuing time into account. This is a quite important notion, since we find that for Android a task is usually queued up in many scenarios of asynchronous execution mechanisms. For example, a task may be waiting for a free thread in a busy thread pool, or waiting for its corresponding handler to be free (details for such mechanisms will be discussed later).The queue time also contributes to how fast the intended task is completed, and as a result, influences the user experiences on UI performance. Our field experiences have shown that many performance problems are caused by a long queuing time of the tasks, which will be shown in our experimental study.

Knowing the execution latency and the queuing time of each asynchronous task in various scenarios can greatly help developers understand how the target app performs in practice, and locate potential performance problems. But the challenge in profiling such execution latency and the queuing time lies in the fact that there are tremendous way to start asynchronous tasks. There are no sole entry and exit for diverse types of asynchronous executions. As a result, it is difficult to find a holistic treatment in tracking the asynchronous tasks. In what follows, we will present how we attack this challenge with a careful taxonomy of the asynchronous tasks and classify them into six categories (in Section \ref{paper_android_perf_icse: profile-taxonomy}). Then we can specifically profile the execution latency of the tasks for each category (discussed in Section \ref{paper_android_perf_icse: profile-implementation}).

\subsection{Android Asynchronous Executions: A Taxonomy}
 \label{paper_android_perf_icse: profile-taxonomy}

As we have discussed, there are hundreds of ways for developers to implicitly or explicitly start asynchronous executions. It is difficult, if not infeasible, to design specific profiling mechanism for each.

If we consider the asynchronous executions started by the \main thread, there could be only two ways: 1. create a new thread 2. reuse pre-created threads.  For the later one, it could be further divided into two types: 2.1. obtain an existing thread directly 2.2. use an existing thread indirectly by a delegate.  Via reviewing various of asynchronous execution methods in Android, we find the mapping of 1, 2.1, 2.2 as separately {\em Looper \& Handler} mechanism, {\em Pool-based Executor} mechanism, and {\em New Thread} mechanism. For every mechanism, there are several representative classes served as the basic class of that type.  Hence, we summarize $6$ basic classes of the mechanisms that an app can start an asynchronous task, which are listed in Table \ref{tab:perf_test_asycategory}.

%初步按照使用方式分类
The $6$ basic classes resort to three basic types of asynchronous execution mechanism of Android, specifically, 1) {\em Looper \& Handler} mechanism, 2) {\em Pool-based Executor} mechanism, and 3) {\em New Thread} mechanism.

\textit{HandlerThread}, \textit{IntentService} and \textit{AsyncQueryHandler} all depend on the Looper \& Handler mechanism to start asynchronous tasks. They create a {\em worker thread} and waiting for new tasks to come on a {\em looper} associated with the worker thread. The request of starting an asynchronous task is sent via a {\em handler} attached to the looper. And then the requested task would be processed on the worker thread. Since there is only one thread, only one message could be processed at one time.  The other requests are queuing in the looper.

\textit{ThreadPoolExecutor} and \textit{AsyncTask} both use the pool-based executor mechanism.  They maintain a thread pool. The number of threads in the pool could be dynamically grow and shrink within a pre-defined range. When the request of starting a asynchronous task comes, the task will be executed in one of the threads in the pool, given that the thread is free ({\em i.e.}, not running other task currently). When the thread pool is full, the new coming tasks have to waiting until an existing task finishes its execution.

Finally, the new-thread mechanism is relatively simple. Its building basis, {\em i.e.}, the \textit{Thread} class in Android, is the same with the traditional Java one. It executes the requested task immediately by starting a new thread.

To summarize, we can see that these six categories of asynchronous executions all rely on two threading mechanisms. The first mechanism creates and allows each asynchronous task to run on a new thread every time. The second creates one or more threads in advance, and wait for tasks to come. The new-arrival asynchronous tasks run on these threads. In other words, threads can be reused other than running for each single task only. The Looper \& Handler mechanism and the Pool-based Executor mechanism belong to the latter.

\begin{table}[!t]
\renewcommand{\arraystretch}{1.3}
\caption{Asynchronous execution categories}
\label{tab:perf_test_asycategory}
\centering
\begin{tabular}{|l||l||l|}
\hline
Category & Type & Representative class\\
\hline
\hline
Non-reusable threading& Independent Thread & Thread\\
\hline
\multirow{5}{*}{Reusable threading} & \multirow{3}{*}{Looper \& Handler} & HandlerThread\\
\cline{3-3}
&& IntentService\\
\cline{3-3}
&& AsyncQueryHandler\\
\cline{2-3}
& \multirow{2}{*}{Pool-based executor} & ThreadPoolExecutor\\
\cline{3-3}
&& AsyncTask\\
\hline
\end{tabular}
\end{table}

\subsection{Profiling Mechanisms}
 \label{paper_android_perf_icse: profile-implementation}

%AsyncTask和Intentservice如何关联相应的部分
%AsyncTask的几个时间点如何采集

We now starts to discuss how track each categories of asynchronous tasks one by one.

1). \texttt{Thread}: An asynchronous task that implements as a thread always starts
the \texttt{start} method of \texttt{Thread}. Hence, we can instantly obtain the
request time of the task ({\em i.e.}, when it is scheduled) when the \texttt{start} method is invoked.
The call-stack can also be saved at the same time as the execution context.

The task is performed in the concrete, overridden \texttt{run} method of a \texttt{Runnable} object.
Although the developer can name a \texttt{Runnable} class arbitrarily, we can resort to a static analysis
approach (shown in Figure \ref{fig:perf_test_framework}) to find such \texttt{run} methods. The information
can be provided to the profiler so that it can track and obtain the execution latency of the method, which
is also that of the task.

2). \texttt{HandlerThread}: \texttt{HandlerThread} is a thread that provides a \texttt{Looper} attached to it.
A \texttt{Handler} is associated to the \texttt{Looper} object and handles message for the \texttt{Looper} (which
is essentially for the thread of \texttt{HandlerThread}). Hence, we can track the request time of a task by tracking when a \texttt{Message} is sent to the \texttt{Handler},
as well as the execution context. Although there are many ways to send a \texttt{Message}. Eventually,  \texttt{sendMessageAtTime} or
\texttt{sendMessageAtFrontOfQueue} must be invoked. Hence, we can track the executions of these two methods
in order to track the \texttt{Message} sending.

\texttt{Handler} actually conducts the task by processing its corresponding \texttt{Message}, which can be tracked
by instrumenting its \texttt{dispatchMessage} methods. In this way, we can obtain queuing time and the execution latency
of the task.

3). \texttt{AsyncQueryHandler}: \texttt{AsyncQueryHandler} is an extending of \texttt{Handler}, which is
widely used in Android coding practice. We use the similar mechanism as that for \texttt{HandlerThread} to obtain its
execution context, the queuing time, and the execution latency.

4). \texttt{ThreadPool}: A \texttt{ThreadPool} task relies on the \texttt{ThreadPoolExecutor} class to run.
The class has an elegant pattern.  A task is always requested
by \texttt{execute} method.  Then it starts with its \texttt{beforeExecute} method called and ends with its
\texttt{afterExecute} method called.  Hence, we can track \texttt{execute} method for obtaining the execution context
and the request time; and the latter two methods for obtaining the queuing time and execution latency eventually.

5). \texttt{Asynctask}: An asynchronous task that relies on the \texttt{AsyncTask} mechanism can base its
implementation on the complicated Java class inheritance of the basis \texttt{AsyncTask} class.
Luckily, we find that no matter how many layers of class inheritance are applied, an asynchronous
task is always eventually scheduled by the \texttt{execute} or \texttt{executeOnExecutor} method
of the \texttt{AsyncTask} class. Hence, we can obtain the request time of the task by tracking the two methods.
The call-stack is also obtained at this moment to save as the execution context.

By inspecting the source codes of \texttt{AsyncTask}, we find that for both cases, \texttt{AsyncTask} actually relies on the thread pool mechanism
to execute the task, which is based on the \texttt{ThreadPoolExecutor} class. We can then use the similar mechanism
as that for \texttt{ThreadPool} to track the required runtime data.

6). \texttt{IntentService}: Each \texttt{IntentService} task always starts by invoking
the \texttt{startService} method of the framework class \texttt{android.app.ContextImpl}.
Hence, we instrument this method for tracking the request time of such tasks, and
save its call-stack as the execution context.

\texttt{IntentService} actually relies on a \texttt{Hanlder} to process the task, by an extended
class \texttt{ServiceHandler} of \texttt{Hanlder}. Hence, we can obtain the time when the task starts
to run and when it terminates by tracking the \texttt{dispatchMessage} method of \texttt{Hanlder}.

Finally, Android is a multi-tasking system that supports concurrent executions of multiple tasks.
The aforementioned methods for each categories of tasks do not always appear sequentially for each
particular task.
For example, the \texttt{beforeExecute} and \texttt{afterExecute} methods of a task can
be interleaved by some \texttt{afterExecute} methods of other tasks. To correctly associate the methods for one particular task is not easy.
In the above design, we resort to different identification approaches to associate the methods for one particular task.

Specifically, for the \texttt{Thread} and \texttt{ThreadPool} tasks, we find that a unique \texttt{Runnable} object is
always a parameter for each method of interest. We then use its hash code as an identifier of a task.
For the \texttt{HandlerThread} and \texttt{AsyncQueryHandler} thread, we find that a \texttt{Message} object is
transmitted along the entire life time of such tasks. Therefore the hash code of \texttt{Message} object is used as the
identifier of a task. For \texttt{AsyncTask} tasks, we observe that the \texttt{AsyncTask} class include the task content
as its private member variable \texttt{mFuture}, which is a \texttt{Runnable} object. This object
remains unchanged during the lifetime of the task. Hence, we adopt its hash code as
an identifier for the task, and accordingly associate the key methods. For the \texttt{IntentService} case, there are unfortunately no global objects that can be used as the identifier.
But, we find that the name of the class extending \texttt{IntentService} is unique for each task. We then
obtain the name as the identifier with the reflection mechanism of Java language.
The implementation of the above association mechanism, although looks simple, actually contains a lot of details to consider. They are
omitted here due to space limitation, which can be found by inspecting the source codes of \texttt{PersisDroid}.

\subsection{Implementation Highlights}

We now discuss how we track the invocations of the methods of interest. We rely on a dynamic
instrumentation mechanism. It require no changes to the target app {\em per se}. It also does not
need to recompile the underlying OS and the Android framework. These features guarantee the compatibility
of \texttt{PersisDroid} when applying in diverse smartphones from
various manufacturers. Moreover, it also requires little human effort in installing and applying the tool.

In our above discussions, we have shown that we specifically tailor the design of \texttt{PersisDroid} so that
the methods it should track are all Android framework methods in Java.
This is first for ease the tracking of these methods, comparing with tracking the functions in OS level which typically
requires heavy-weighted and sophisticated tool for kernel instrumentation. But most importantly, we can thus rely on an Android specific feature to conveniently track these methods.

In particular, Android processes for apps, unlike general Linux processes, are all created by duplicating a system process called \texttt{Zygote}. Framework binaries have already been loaded in \texttt{Zygote} before such duplication. Therefore, we can instrument the \texttt{Zygote}
process and ``hijack'' the framework methods of interest before an app runs. Then when it runs by forking
\texttt{Zygote}, the method invocations are inherently hijacked by \texttt{PersisDroid}. Hence, we can easily track
the methods. We implement this idea by adopting a tool called Xposed \cite{AndroidTool:Xposed}, which usually used for prettifying user interface \cite{AndroidTool:XposedModules}. It can substitute the original \texttt{Zygote} process with a hijacked one. We rely on its mechanism and program our own codes to hijack the methods of our interest.

Finally, as we have discussed, we adopt a static analysis
approach (shown in Figure \ref{fig:perf_test_framework}) to find the \texttt{run} methods for tracking
\texttt{Thread} tasks, since our dynamic instrumentation approach requires such information.
We first use a tool called \texttt{apktool} \cite{Android:apktool} to generate the bytecodes from an app in binary form. Such bytecodes are well structured \cite{Android:dalvik_bytecode}. We then use Python to code a parser to analyze the bytecodes to obtain the \texttt{run} methods and their corresponding information of interest.

All the implementation details can be found in the source codes we have released at \cite{AndroidPerf:PersisDroid}.

\section{Detecting and Locating Performance Anomalies}
\label{paper_android_perf_icse: analyzer}
  
The profiler use the default general \texttt{Logcat} \cite{Android:Logcat} logging mechanism released with Android to save and transmit the logs to the PC. 

As discussed, we assign a unique identifier to the methods of each particular asynchronous task. The log analyzer can instantly obtain 
the execution context, the queuing time, and the execution latency 
of each task. 

Since Android will typically starts a lot of management threads for an app, such threads also include asynchronous tasks in nature. But they
are not relevant to the app performance. In this regard, we first filter out such asynchronous tasks. A simple way is to check whether 
they are requested by the \main UI thread or its offspring threads. 
A offspring thread of the \main thread is that created by the \main thread or that created by another offspring thread of the \main thread.

After the feature data of each asynchronous task of interest is 
obtained, we first group the tasks based on their execution context. 
Those with the same execution context are intuitively the instances of the same task. Then, for each group, we can obtain their statistics. 
\texttt{PersisDroid} can then draw a histogram for each group. 
Such histograms can help developers understand the performance of the app by looking into the performance distribution of each group of asynchronous tasks. For example, a high variance of queuing time may indicate an improper task schedule mechanism. 
  
More importantly, \texttt{PersisDroid} also reports suspicious data where a performance problem may manifest. \texttt{PersisDroid} resorts 
to several heuristics to detect and report such suspicious data. 
Simple examples include that a large variance of the execution latency or waiting time will signal a warning, and that a large deviation between the maximum value and the minimum value of the time can also signal a warning. To avoid the noise caused by the very small but correct execution time 
({\em e.g.}, a method may return immediately when the parameter is invalid, which will result in a very short execution time), we can also 
consider the difference between the maximum value and the medium value as an indicator of suspiciousness. In our experimental study, we will illustrate how the log analyzer help us in pinpointing tens of bugs of the real-world apps in practice.

\section{Experiment}
\label{sec:perf_test_experiment}

We first test the overhead of \texttt{PersisDroid} using the Android \texttt{time} command.  We perform $10000$ monkey operations on our three devices with \texttt{PersisDroid} on and off with operation interval equal to $200$ms. The results shows the average overhead is $0.8\%$ in testing time. This shows that our tool does not have a consideration impact on testing efficiency.

To demonstrate the effectiveness of \texttt{PersisDroid}, we first show an example report it generates, followed by providing our
experiences of using the tool to locate tens of real-world bugs.

\subsection{An Example Report}
\label{subsec:perf_test_reportexample}

\begin{figure*}[!t]
\centering
\includegraphics[width=0.8\textwidth]{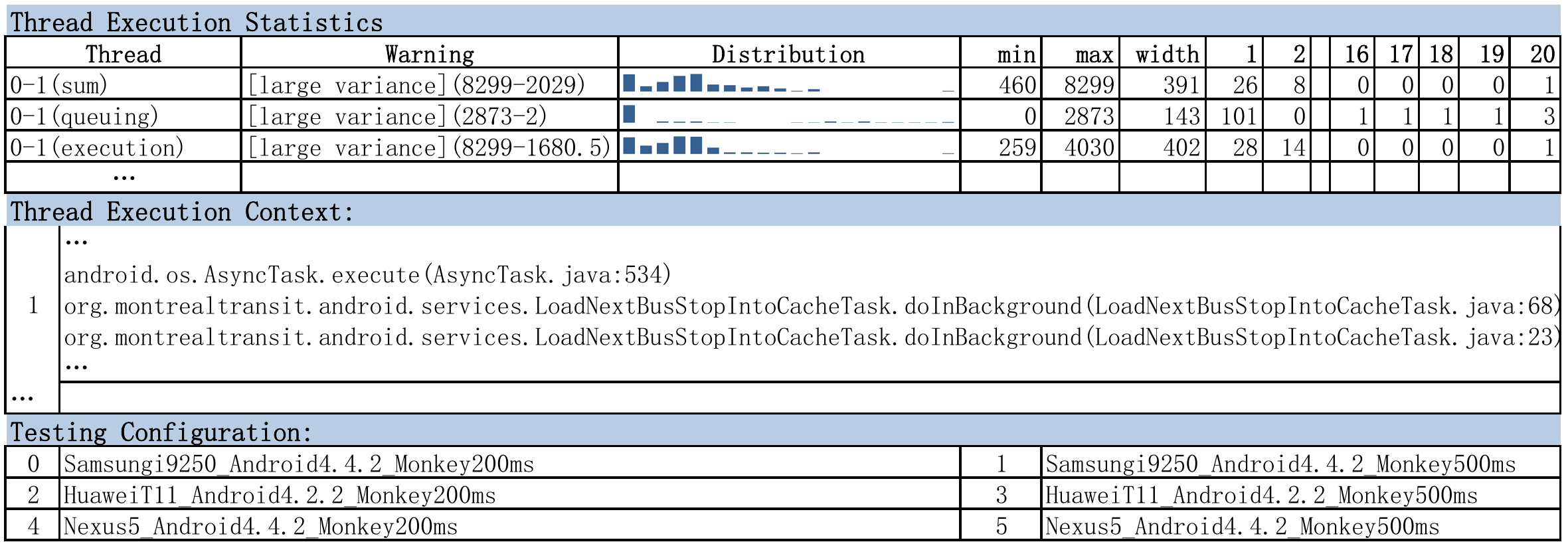}%{perf_test_framework}
\caption{Report example}
\label{fig:perf_test_report}
\end{figure*}

Figure \ref{fig:perf_test_report} shows a simplified report generated by \texttt{PersisDroid}. A report includes three parts
the performance statistics, the execution context, and the testing configurations.
The thread ids, for example ``0-1'' in the figure, means the line of data is generated under testing configuration ``0'' with execution context index ``1''. The corresponding testing configuration and the execution context can be found in the report.

The lines are arranged according to the suspiciousness of the task group. When a warning is signaled, developers can quickly
inspect their codes to examine whether there are defects. In this example case, we can see that the task is an \texttt{AsyncTask}, with warnings on both queuing and execution time.  Developers can inspect LoadNextBusStopIntoCacheTask.java to see whether these are correct behaviors. Finally, \texttt{PersisDroid} allows multiple test runs with different hardware and software configurations. In this example, configuration ``0'' means we test on Nexus5 with OS version 4.4.2, and the interval of the random test operations (with Monkey) is 500ms.

\subsection{Finding real world performance bugs}
With careful compatibility consideration in \texttt{PersisDroid} design, we can easily apply it in different smartphones. In our experiments,
we use three models: Huawei G610-T11 with Android 4.2.2, Nexus5 with Android 4.4.2, and Samsung i9250 with Android 4.4.2.

We download $30$ apps randomly selected from F-Droid \cite{Android:FDroid}, an open-source app market, and apply \texttt{PersisDroid}
to these apps. Surprisingly, we find $20$ contains potential performance issues.  We categorize the detected issues into $6$ categories in Table \ref{tab:perf_test_issues}, together with their causes and their locations.

\begin{table*}[!t]
\renewcommand{\arraystretch}{1.3}
% if using array.sty, it might be a good idea to tweak the value of
% \extrarowheight as needed to properly center the text within the cells
\caption{Representative performance issues found}
\label{tab:perf_test_issues}
\centering
\begin{tabular}{|l||l|l|l|l|}
\hline
Category & Issue description & Class & App & Source\\
\hline
\hline
\multirow{2}{*}{\tabincell{l}{Blocking\\ execution}} & \tabincell{l}{Not responding to any operations when performing\\ asynchronous search} & MainActivity & RestC & Github\\
\cline{2-5}
& Blocking for results of asynchronous execution & LocationAdapter & Liberario & Github\\
\hline
\multirow{4}{*}{\tabincell{l}{Not canceling\\ execution}} & Do not cancel tuning tasks when pages switched off & LibreDroid & GNU FM & Gitorious\\
\cline{2-5}
& Cannot cancel loading a JSON file & \tabincell{l}{LoadNextBusStop-\\IntoCacheTask} & MonTransit & Google code\\
\cline{2-5}
& \tabincell{l}{Could only cancel before but not interrupt downloading} & \tabincell{l}{DownloadImages-\\Task} &  \tabincell{l}{Helsinki Testbed\\ Viewer 2.0} & Github\\
\cline{2-5}
& Do not interrupt the network related tasks & PlatformUtil & PageTurner & Github\\
\hline
\multirow{2}{*}{\tabincell{l}{Execution\\ pool}} & \tabincell{l}{All tasks using the public pool\\ \texttt{AsyncTask.THREAD\_POOL\_EXECUTOR} occupy the pool} & PlatformUtil & PageTurner & Github\\
\cline{2-5}
& \tabincell{l}{The developer does not know the optimal size of thread\\ pool and writes (how many threads ???) in remarks} & \tabincell{l}{ZLAndroidImage-\\Loader} & FBReader & Github\\
\hline
\tabincell{l}{Message\\ queue} & \tabincell{l}{Use the same message queue of a background thread makes\\ large queuing time} & ReadingFragment & PageTurner & Github\\
\hline
\multirow{5}{*}{\tabincell{l}{Sequential\\ execution}} & \tabincell{l}{Always wait for the last operation in a separate thread\\ to finish} & \tabincell{l}{IntermediatePoints-\\Dialog} & OsmAnd & Github\\
\cline{2-5}
& Long-term operation queuing serially & SubwayLineInfo & MonTransit & Google code\\
\cline{2-5}
& Use serial pool for all tasks & LibreDroid & GNU FM & Gitorious\\
\cline{2-5}
& \tabincell{l}{Remarking parallel executing but use \texttt{execute}\\ function of \AsyncTask which is serial} & LawListFragment & OpenLaw & Github\\
\cline{2-5}
& Respondings to clicks are serial & \tabincell{l}{AnimateDraggingMap-\\Thread} & OsmAnd & Github\\
\hline
\multirow{2}{*}{\tabincell{l}{Third-party\\ library}} & \tabincell{l}{The library implements cancel operation for long-term\\tasks, but not called by the app} & \tabincell{l}{HeadlineComposer-\\Adapter} & OpenLaw & Github\\
\cline{2-5}
& \tabincell{l}{Problematic implementation of \texttt{Filter} makes response\\ laggy for inputting in \texttt{AutoCompleteTextView}} & LocationAdapter & Liberario & Github\\
\hline
\end{tabular}
\end{table*}

1). \texttt{Blocking execution}: Blocking execution is a type of bugs that incur large execution time, and consequently block the UI.
A representative example is the RestC app. \texttt{PersisDroid} detects that \texttt{RetrieveFeedTask}
is time-consuming. But the UI, after showing a \texttt{ProgressDialog}, is busy waiting for the task to be done.
But task performs Internet access. When network condition is poor, UI will be long blocked.

2). \texttt{Not canceling execution}: If an asynchronous task should not be running when it is no longer useful.
\texttt{PersisDroid} reports that several apps contains such bugs.
For example, the \texttt{DownloadImagesTask} of Helsinki Testbed Viewer 2.0 can although check whether to cancel itself
before the image is really downloading, it does not perform such checks when the downloading starts. This can cause expected
execution latency, which can be instantly captured by \texttt{PersisDroid}.

3). \texttt{Execution pool}: A typical misuse of thread pool is overload, which incurs long queuing time since the pool is usually busy.
\texttt{PersisDroid} can easily detect such anomalies. For example, PageTurner project typically uses the \texttt{PlatformUtil} class in starting \texttt{AsyncTask}s, which are always executed on the same pool. Since there are many \texttt{AsyncTask} executions, the pool
is always busy, causing a long queuing time.

4). \texttt{Message queue}: Message queue is similar to thread pool. But there are usually multiple worker threads run in a thread pool, but only one worker thread being attached with the message queue. Overloading is a quite common bug. For example, in \texttt{ReadingFragment} of PageTurner project, a handler is attached to a background \texttt{HandlerThread}. The handler is quite busy but there is only one background thread
to process its request, which inevitably causes a large queuing time. This can also easily be pinpointed with  \texttt{PersisDroid}.

5). \texttt{Sequential execution}: Many developers tend to use parallel execution to improve efficiency.  However, misuse of parallel execution may cause the execution become serial.  We detect the sequential execution by the indicator of queuing time.  Serial executions always own a large queuing time since the probability of getting conflict is high.  By looking into the related methods and classes that have a large queuing time.  A common mistake is use \texttt{execute} method instead of \texttt{executeOnExecutor} method of \texttt{AsyncTask}.  It is not hard to detect this kind of cases.

%We would like to show a more complex example on sequential threading.  \texttt{AnimateDraggingMapThread} is the utility class that %performs UI operations and animations like moving, dragging and zooming of OsmAnd project.  For every animation it invokes %%%\texttt{startThreadAnimating} method.  Latency is found in this method by our report.  Looking into this method, we find it would first %block and wait for the last move operation to be processed.  In this case, it cannot process multiple UI inputs efficiently.  A suggestion %is that the developer could accept multiple UI inputs and then combine the unprocessed ones into a single UI input.

6). \texttt{Third-party library}: Misusing of the third-party libraries is also the source of bugs. \texttt{PersisDroid} can also
find suspicious executing time and queuing time in third-party libraries.  For example, the \texttt{HeadlineComposerAdapter} object in the OpenLaw project uses a third-party API \texttt{AsynHttpClient} to get http responses asynchronously. When the object
is no longer needed, the task in the API should also be cancelled. But the developers mistakenly
code the canceling mechanism and never actually cancel such tasks as a result, which can be instantly located by \texttt{PersisDroid}.

\section{Related Work}
\subsection{Android application testing}
Testing of Android app is widely studied in both industry and academic.  Currently, referring to automatic testing of Android application, the common practice in industry is script testing.  To name a few, UIAutomator \cite{AndroidTest:UIAutomator} as well as Monkey runner \cite{AndroidTest:MonkeyRunner} are both packed together with official development tools; Robotium \cite{AndroidTest:Robotium} is a representative third party script testing tool.  However, it takes time to write scripts in high quality.  More than that, the scripts cover only limited user scenarios.

Researchers resort to fully-automatic testing as a complementary for script testing.  Many works are done to improve the test coverage and efficiency.  The approaches generally lie in three categories, namely fuzz testing, systematic exploring and model-based testing.  Fuzz testing approaches generate random input sequences for testing Android apps.  Examples of fuzz testing include Monkey \cite{AndroidTest:Monkey}, Dynodroid \cite{AndroidTest:Dynodroid} and VanarSena \cite{AndroidTest:CloudMonkeyTest} as a scalable solution.  Symbolic execution base testing aims at exploring the app functions in a systematical way.  Mirzaei et al. \cite{AndroidTest:SymoblicExec}, ACTEve \cite{AndroidTest:Auto_conclolic}, Jensen et al. \cite{AndroidTest:TargetSeqGen} and $A^3E$ \cite{AndroidTest:TargetedDFS} are the representatives.  Model-based testing aims at generating a finite state machine model and event sequences to traverse the model.  Examples of this category include Android Ripper \cite{AndroidTest:AndroidGUIRipping}, Android GUITAR \cite{AndroidTest:AndroidGUITAR}, MobiGUITAR \cite{AndroidTest:MobiGUITAR} and SwiftHand \cite{AndroidTest:SwiftHand}.

However, all these works focus on functional testing which mainly test the functionality of an app.  They are not designed for performance testing.

\subsection{Performance measuring for smartphone}
Performance testing, diagnosing and improving is also studied on mobiles.  Amrutkar et al. \cite{AndroidPerf:MobilenetDiagnosis} diagnosis the under-performance of mobile Internet.  Huang et al. \cite{AndroidPerf:NetworkAppDiagnosis} study the performance of network applications.  Timecard \cite{AndroidPerf:ServerbasedApp} aims at controlling the delay between server-based mobile apps.  Toolder \cite{AndroidPerf:MemoryPatternDetect} detects performance degradation caused by a specialized memory access pattern.  Kim et al. \cite{AndroidPerf:Storageperformance} benchmarks the storage performance of Android devices.  Thompson et al. \cite{AndroidPerf:ResourceConsumModel} apply a model based estimation of app performance in terms of resource consumption.   Wang et al. \cite{AndroidPerf:OffloadingModeling} use a Stochastic Activity Network model for the performance of the offloading tasks for mobile in terms of unstability, throughput and energy.  However, the definitions of performance are in these works are not UI related performance as in our work.

Liu et al. \cite{AndroidPerf:CharacterizingBugs} first characterize the UI related performance issues.  They find that many performance issues are caused by blocking operations in the \main thread.  StrictMode \cite{AndroidPerf:StrictMode} and Asynchronizer \cite{AndroidPerf:RefactorAsynctask} have done the work of analyze the \main thread for blocking operations.

There are also works for general purposed UI performance diagnosis.  Appinsight \cite{AndroidPerf:Appinsight} is the first tracing based performance testing tool on mobile apps.  It traces all related executions on any threads (including both \main thread and worker threads) from when event triggered to the corresponding UI update.  The routine traced is used as the performance of the application.  Appinsight is designed for Windows phone.  Panappticon \cite{AndroidPerf:Panappticon} follows a similar idea on Android platform.  However there are several drawbacks of this solution. 1. Complicated:  the tool records too much information which makes it hard to install and use 2. Compatibility: Panappticon relies highly on the injection of the framework and also the kernel which makes it not easy to be applied on new Android os versions or different devices 3. No queuing time: as discussed in Section \ref{sec:perf_test_experiment} there are many performance bugs make the queuing time longer.  Queuing time is a very important indicator.

%\section{Discussion}
%\input{perf_test_discussion}

\section{Conclusion and future work}
%这篇paper的总结，我们先讨论异步任务重要性，然后对异步任务分类并profile，我们讲profiler整合到性能测试的框架中，并实现性能测试工具。最后我们展示我们的工具确实能有效的找到性能问题，且给开发者提供一些建议。
In this paper, we discussed performance testing of Android applications.  We first addressed a sort of important performance issues that are caused by misuse of asynchronous executions.  Some motivating examples were given to show the misusage.  Then we classified categories of asynchronous executions through inspecting Android framework source code.  A profiler is implemented for monitoring the performance of asynchronous executions during run time.  After that we introduced \texttt{PersisDroid}, the implementation of a performance testing framework for Android apps.  The profiler is integrated into \texttt{PersisDroid}.  \texttt{PersisDroid} would generate a visualized report containing the distribution of asynchronous executions of the app.  Finally we conduct extensive experiments to show the powerfulness of \texttt{PersisDroid}.  We illustrated the usage of our report and show how real performance bugs are found with assisting of the report.

%\subsection{Tool overhead}
%We leave the task of optimizing overhead of this tool for future work as we believe this is not a critical issue for the testing phase.  Moreover, our tool could be used for the initial performance test only.  That means our tool would find some suspicious asynchronous executions then the developers could do the method hooking themselves.
Currently there are still some limitations on \texttt{PersisDroid}.  The first is that we do not test problems on \main thread. Those performance issues caused by bad design of \main thread will not be detected by our tools.  The second is that \texttt{PersisDroid} only provide a visualized report to the developers in this time.  The performance tuning still relies on the experience of the developers.  Therefore the possible future work is making the performance tuning fully automatical and including analysis of \main thread as well.

\bibliographystyle{IEEEtran}
\bibliography{../bib/Android_performance,../bib/Android,../bib/Android_test}
\end{document}